\documentclass [12pt,a4paper]{article}
\usepackage{amssymb, theorem}
\usepackage{graphics}
\usepackage{graphicx}
\def\bbbc{{\mathchoice {\setbox0=\hbox{$\displaystyle\rm C$}\hbox{\hbox
to0pt{\kern0.4\wd0\vrule height0.9\ht0\hss}\box0}}
{\setbox0=\hbox{$\textstyle\rm C$}\hbox{\hbox
to0pt{\kern0.4\wd0\vrule height0.9\ht0\hss}\box0}}
{\setbox0=\hbox{$\scriptstyle\rm C$}\hbox{\hbox
to0pt{\kern0.4\wd0\vrule height0.9\ht0\hss}\box0}}
{\setbox0=\hbox{$\scriptscriptstyle\rm C$}\hbox{\hbox
to0pt{\kern0.4\wd0\vrule height0.9\ht0\hss}\box0}}}}

\def\qed{{\rightline \sq}\medskip}
\newtheorem{thm}{Theorem}
\newtheorem{prop}{Proposition}

\theorembodyfont{\rm}
\newtheorem{pl}{Example}
\newtheorem{example}{Example}

\renewcommand{\qed}{\nobreak\hfill $\square$\medskip}
\def\proof{\noindent{\it Proof.} }
\def\<{\langle}
\def\>{\rangle}
\def\fel{\textstyle{1 \over 2}}
\def\eps{\varepsilon}
\def\iH{{\cal H}}
\def\iM{{\cal M}}

\def\iD{{\cal D}}
\def\iX{{\cal X}}
\def\bA{{\mathbf A}}

\def\bu{{\bold u}}
\def\b1{{\bf 1}}

\def\eps{\varepsilon}

\def\Tr{{\rm Tr}}

\def\im{{\rm i}}

\def\Tr{\mbox{Tr}\,}
\def\Prob{\mbox{Prob}}
\def\Diag{\mbox{Diag}}
\def\Det{\mbox{Det}}

\def\bbbr{{\mathbb R}}

\def\argmin{{\rm argmin}}
\def\rea{{\rm Re}\,}
\def\ima{{\rm Im}\,}
\def\treff{\medskip $\clubsuit \qquad$}
\def\un{{\rm un}}
\begin{document}
\vspace{-2cm}
%\rightline{\today}
\ \vskip 1cm
\centerline{\LARGE {\bf Point Estimation of States}}
\bigskip
\centerline{\LARGE {\bf of Finite Quantum Systems}}
\bigskip
\medskip\footnotetext[3]{Supported by the Hungarian Research Grants OTKA
T042710 and T032662.}
\bigskip
\centerline{ D. Petz$^{1,3}$, K.M. Hangos$^{2,3}$  and A. Magyar$^{2,3}$}
\medskip
\bigskip
\begin{center}
$^1$ Alfr\'ed R\'enyi Institute of Mathematics, \\ H-1364 Budapest,
POB 127, Hungary
\end{center}
%\medskip
\begin{center}
$^2$ Computer and Automation Research Institute, \\ H-1518 Budapest, POB 63,
Hungary
\end{center}
\medskip
\bigskip
\begin{quote}
{\bf Abstract:} The estimation of the density matrix of a $k$-level quantum system
is studied when the parametrization is given by the real and imaginary part of
the entries and they are estimated by independent measurements. It is established
that the properties of the estimation procedure depend very much on the invertibility
of the true state. In particular, in case of a pure state the estimation is less
efficient. Moreover, several estimation schemes are compared for the unknown state
of a qubit when one copy is measured at a time.
It is shown that the average mean quadratic error matrix is the smallest
 if the applied observables are complementary.
The results are illustrated by computer simulations.



\end{quote}

\section{Introduction}
The problem of inferring the state of a quantum system from measurement data is
fundamental. One side of this problem is the adequate experimental techniques,
and the other side is the theory based on the adaptation of statistics to the
quantum mechanical formalism.

Most of the work in state estimation has focused on states of a qubit, pure
states \cite{GM}, or mixed states \cite{Bagan, We, REK}. The estimation 
procedure for pure states is simpler, partially due to the smaller number
of parameters. The subject of the present paper is state estimation for
a $k$-level quantum system. In this case the boundary of the state space
is not the set of pure states but the non-invertible density matrices. The
entries of the density matrix provide a natural parametrization of the 
state space. The accuracy of the estimation can be quantified  by the 
fidelity or by the Hilbert-Schmidt distance. For larger matrices the latter 
seems to be easier to handle. When different estimation schemes are compared, 
the mean quadratic error matrix can be used. 

\section{The estimation scheme}

The goal of  state estimation is to determine the density operator $\rho$
of a quantum system by measurements on $n$ copies of the quantum system
which are all prepared according to $\rho$ \cite{Janos, Dariano, We}. 
The number $n$ corresponds to
the sample size in classical mathematical statistics. An estimation scheme
means a measurement and an estimate for every $n$. For a reasonable
scheme, we expect  the  estimation error to tend to 0 when $n$ tends to
infinity as a consequence of the law of large numbers.

Assume that $\rho$ is the density matrix of our system described on the Hilbert
space $\iH$. Then the $n$ identical copies are described by the $n$-fold tensor
product $\iH_n:=\iH^{n\otimes}$ and the state is $\rho_n:=\rho^{n\otimes}$.
When $\dim \iH=k$, we can identify the operators of  $\iH_n$ with matrices of
$k^n \times k^n$. In this paper we study measurement schemes given by self-adjoint
matrices
\begin{equation}
\bA(n)=(\bA(n)_{ij})_{i,j=1}^k,
\end{equation}
where $\bA(n)_{ij}\in B(\iH_n)$. Note that $\bA(n)$ is determined by $k^2$
self-adjoint operators acting on $\iH_n$. They are the diagonal entries
$Z(n)_{ii}\equiv \bA(n)_{ii}$ of $\bA(n)$, moreover the off-diagonal entries
are written as
$$
\bA(n)_{ij}=X(n)_{ij}+\im Y(n)_{ij} \qquad (i <j)
$$
by means of self-adjoint $X(n)_{ij}$ and $Y(n)_{ij}$. The measurement scheme
$\bA(n)$ means that the observables $Z(n)_{ii}$, $X(n)_{ij}$ and $Y(n)_{ij}$
are measured on the $r$ copies of the original system. Since the sum of the
diagonal entries of a density matrix is 1, it is enough to measure $k-1$
diagonal entries, for example, $Z(n)_{kk}$ can be removed from the set
of observables to be measured and $k^2-1$ observables remain.
(Hence $n=r(k^2-1)$.)

\begin{example} \label{pl:1}
Let $k=2$ and
$$
S_n(\sigma_i)=\frac{1}{n}(\sigma_i \otimes I \otimes \dots \otimes I+
I\otimes \sigma_i \otimes I \otimes \dots \otimes I+ \dots +
I \otimes I \otimes \dots \otimes \sigma_i)\in B(\iH_n),
$$
where $1 \leq i \leq 3$ and $\sigma_i$ are the Pauli matrices. Set
\begin{equation} \label{E:An}
\bA(n) = \frac{1}{2}\left( \begin{array}{cc}
I_n+S_n(\sigma_3)  & S_n(\sigma_1)-\im S_n(\sigma_2)\\
S_n(\sigma_1)+\im S_n(\sigma_2)  & I_n-S_n(\sigma_3)
\end{array} \right) ,
\end{equation}
where $I_n$ denotes the identity on $\iH_n$.

We may have a better understanding of this estimation scheme if
the $n$-fold product is considered to be embedded into the
infinite product. Then the limit $n\to \infty$ is more visible. If
$$
\rho=\left( \begin{array}{cc}
\rho_{11}  & \rho_{12} \\
\rho_{21} & \rho_{22}
\end{array} \right) ,
$$
then the law of large numbers guarantees that
$$
\bA(n)_{ij} \to \rho_{ij} I_\infty,
$$
where $I_\infty$ denotes the identity in the infinite tensor product.
Therefore, the error is going to 0 when $n$ goes to $\infty$ for any
reasonable definition of the error.

The entries of the matrix (\ref{E:An}) do not commute, therefore there is no
joint Kolmogorovian model for them and the observables cannot be measured
simultaneously. We modify this matrix, in order to use standard
probabilistic tools.

On the infinite tensor product $M_k \otimes M_k \otimes \dots$ we introduce
the right shift $\gamma$:
$$
\gamma (H_1 \otimes H_2 \otimes \dots H_n \otimes I_k \otimes I_k \dots )=
I_k \otimes H_1 \otimes H_2 \otimes \dots H_n \otimes I_k \otimes I_k \dots
$$
Now we set
\begin{equation} \label{E:Bn}
\hat\bA(n) = \frac{1}{2}\left( \begin{array}{cc}
I_r+S_r(\sigma_3)  & \gamma^r \big(S_r(\sigma_1)\big)-\im \gamma^{2r}
\big(S_r(\sigma_2)\big)\\
\gamma^r \big(S_r(\sigma_1)\big)+\im \gamma^{2r}\big(S_r(\sigma_2)\big)
& I_r-S_r(\sigma_3)
\end{array} \right) .
\end{equation}
The operators $S_r(\sigma_3)$, $\gamma^r \big(S_r(\sigma_1)\big)$ and
$\gamma^{2r}\big(S_r(\sigma_2)\big)$ commute. They may be regarded as classical
random variables, one can speak about their joint distribution, variance etc.
\qed \end{example}

The very concrete estimation scheme we use will be the natural extension of
Example \ref{pl:1}. Denote by $E_{ij}$ the $k \times k$ matrix units and set
\begin{eqnarray*}
Z_{ii}&:=&\gamma^{\tau(i,i)}(E_{ii}) \qquad \qquad (1 \leq i <k), \cr
X_{ij}&:=&\gamma^{\tau(i,j)}(E_{ij}+E_{ji}) \qquad (i < j),\cr
Y_{ij}&:=&\gamma^{\tau(j,i)}(\im E_{ij}-\im E_{ji})\qquad (i < j),
\end{eqnarray*}
where $\tau:\{(i,j):1\leq i, j \leq k, (i,j)\neq(k,k)\} \to \{1,2,\dots, k^2-1\}$
is an arbitrary bijection. These self-adjoint operators commute and behave as
independent random variables. The spectrum of $Z_{ii}$ is $\{0,1\}$ and the
spectrum of $X_{ij}$ and $Y_{ij}$ is $\{-1,0,1\}$. The matrix $\bA(k^2-1)$ is
determined by these operator entries.

Finally, the estimation scheme $\bA(r(k^2-1))$ is defined by the formulas
\begin{eqnarray*}
Z(r(k^2-1))_{ii}&:=&\frac{1}{r}\sum_{m=0}^{r-1}
\gamma^{m(k^2-1)}(Z_{ii}) \qquad (1 \leq i <k), \cr
X(r(k^2-1))_{ij}&:=&\frac{1}{r}\sum_{m=0}^{r-1}
\gamma^{m(k^2-1)}(X_{ij}) \qquad (i < j), \cr
Y(r(k^2-1))_{ij}&:=&\frac{1}{r}\sum_{m=0}^{r-1}
\gamma^{m(k^2-1)}(Y_{ij}) \qquad (i < j).
\end{eqnarray*}
Note that to carry on the measurement of all these observable $r(k^2-1)$
copies of the original quantum system are needed. The entries of $\bA(n)$
are commuting observables, therefore there is a basis in $\iH_n$ such that
all of them are diagonal in this basis. Consequently, a single measurement
can be performed theoretically instead of the measurements of the $k^2-1$
observables ($n=r(k^2-1)$).

Our aim is to estimate the $k \times k$ density matrix $\rho$ of a quantum
system. The parametrization is naturally given by the entries of the matrix.
In what follows, we are given several copies of a $k$-level quantum system in
the same state. We perform measurements on the systems one after another, that
is, a system is measured only once, the next measurement is performed on
the next copy of the system, so the states of the systems after the
measurement are irrelevant from our viewpoint.

If we want to estimate the real part of $ij$ entry of the density
matrix $\rho$, then we measure the observable  $E_{ij}+E_{ji}$. Its spectral
decomposition is
$$
1\cdot\frac{1}{2}(E_{ii}+E_{ij}+E_{ji}+E_{jj})+ 0\cdot \sum_{i\ne m \ne j} E_{mm}
-1\cdot\frac{1}{2}(E_{ii}-E_{ij}-E_{ji}+E_{jj})
$$
and its measurement has three different outcomes, $\pm 1$ and $0$. The
probabilities of the outcomes $\pm1$ are
$$
\Prob (X_{ij}=\pm 1)  = \frac{1}{2}(\rho_{ii} \pm \rho_{ij} \pm\rho_{ji}+\rho_{jj})=
\frac{1}{2}(\rho_{ii}+\rho_{jj})\pm \rea \rho_{ij} \,.
$$
To estimate the imaginary part, we measure $\im E_{ij}-\im E_{ji}$
with spectral decomposition
$$
1\cdot\frac{1}{2}(E_{ii}+\im E_{ij}-\im E_{ji}+E_{jj})+ 0\cdot \sum_{i\ne m \ne j}
E_{mm} -1\cdot\frac{1}{2}(E_{ii}-\im E_{ij}+ \im E_{ji}+E_{jj}).
$$
The probabilities are
$$
\Prob (Y_{ij}=\pm 1)  = \frac{1}{2}(\rho_{ii} \pm \im \rho_{ij} \mp \im \rho_{ji}
+\rho_{jj})= \frac{1}{2}(\rho_{ii}+\rho_{jj})\pm \im \ima \rho_{ij} \,.
$$
Finally, for the diagonal $ii$ entry we have
$$
\Prob (Z_{ii}= 1)  = \rho_{ii}.
$$

All the three kinds of measurement are performed $r$ times. If $M$ is one
of the measurements which has outcome $t$, the we denote by $\nu(r, M, t)$
the relative frequency of $t$ when the measurement is performed $r$ times.
According to the {\bf law of large numbers}, $\nu(r, M, t)\to \Prob(M=t)$ as
$r \to \infty$. The following estimate is natural:
\begin{itemize}
\item [(i)]
$\Phi_n^{\un}(\rho)_{ii}=\nu(r, Z_{ii}, 1)$ for $(1 \leq i <k)$ and
$$
\Phi_n^{\un}(\rho)_{kk}=1-\sum_{i=1}^{k-1}\nu(r, Z_{ii}), ,
$$
\item [(ii)]
$\rea \Phi_n^{\un}(\rho)_{ij}=\fel (\nu(r, X_{ij}, 1)-\nu(r, X_{ij}, -1))$ for
$i < j$.
\item [(iii)]
$\ima \Phi_n^{\un}(\rho)_{ij}=\fel (\nu(r, Y_{ij}, 1)-\nu(r, Y_{ij}, -1))$
for $i < j$.
\end{itemize}

In our notation, ``{\it un}'' is an abbreviation of the word ``{\it
unconstrained}''. It may happen that $\Phi_n^{\un}(\rho)$ is not a positive
semidefinite matrix, hence it is not an estimate in the really strict sense.
Let $\iM_k$ denote the set of all self-adjoint $k \times k$ matrices of trace
1. The estimate $\Phi_n^{\un}$ takes its values in $\iM_k$. Note that the set
of invertible density matrices form an open subset of $\iM_k$.

Given a true state $\rho$, $\Phi_n^{\un}$ is a matrix-(or vector-)valued
random variable which is the mean of $r$ independent copies of $\Phi_0^{\un}$.
Let $G\subset \iM_k$ be an open set such that $\rho \in G$. According to
the law of large numbers
$$
\Prob (\Phi_n^{\un} \notin G) \to 0,
$$
however according to the {\bf large deviation theorem} the convergence
is exponentially fast:
$$
\Prob (\Phi_n^{\un} \notin G) \le C \exp (-n E_G),
$$
where $E_G>0$ is the infimum of the so-called rate function, see
\cite{Ellis}.

\begin{thm}
Assume that $\rho$ is an invertible density matrix. The probability of that
$\Phi_n^{\un}$ is not a density matrix converges exponentially to 0
as $n \to \infty$.
\end{thm}

\proof
The expectation value of $\Phi_1^{\un}$ is $\rho \in \iM_k$. Cram\'er's
theorem tells us that there is a function $I:\iM_k \to \bbbr^+\cup \{+\infty\}$
such that for any open set containg $\rho$
$$
\limsup_{n \to \infty}\frac{1}{n}\log \Prob ( \Phi_n^{\un} \notin G)
\le  -\inf \{ I(D): D \in \iM_n \setminus G\}
$$
The RHS is strictly negative and if $\rho$ is invertible, then we can choose
$G$ such that it consists of density matrices (that is, its elements are positive
definite). This gives the proof.

The computation of the rate function $I$ is theoretically possible, but we do not
need its concrete form. \qed

Although the expectation value of the unconstrained estimate $\Phi_n^{\un}$
is the true state, this does not mean that $\Phi_n^{\un}$ is a good estimate.
It may happen that the value of $\Phi_n^{\un}$ is outside of the state
space with some probability.

\begin{figure}[!ht]
\begin{center}
\includegraphics[width=14cm]{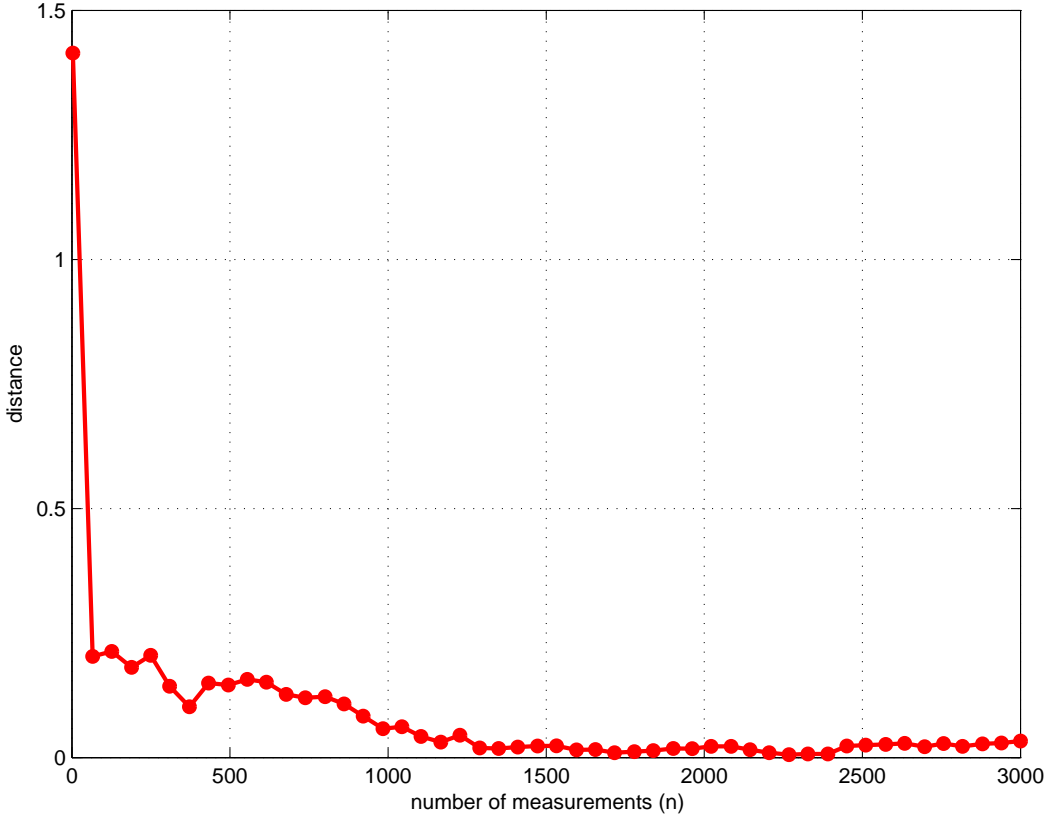}
\end{center}
\caption{\label{F:dist_2}
The Hilbert-Schmidt distance between the true pure state
(\ref{E:pure}) and $\Phi_n^{\un}$ does not converge to 0 as $n \to \infty$.}
\end{figure}

\begin{pl}
Consider the pure state
\begin{equation} \label{E:pure}
\rho=\frac{1}{2}\left[ \begin{array}{cc} 1&1\\ 1&1\end{array} \right]=
\frac{1}{2}(\sigma_0+\sigma_1).
\end{equation}
Then $Z_{11}$ is a random variable $\eta_1$ such that $\Prob (\eta_1=1)=
\Prob (\eta_1=0)=1/2$, $X_{12}$ is 1 (with probability 1) and  $Y_{12}$
is a random variable $\eta_2$ such that  $\Prob (\eta_2=\pm 1)=1/2$.

One can compute that the expectation value of the determinant of
$\Phi_n^{\un}$ equals $-3/2$, independently of $n$. Therefore, in this
example $\Phi_n^{\un}$ is a rather bad estimate, for example, it is not
true that the probability of indefinite estimate goes to 0 as $n \to \infty$,
see also Figure \ref{F:dist_2}.\qed
\end{pl}

\begin{figure}[!ht]
\begin{center}
\includegraphics[width=14cm]{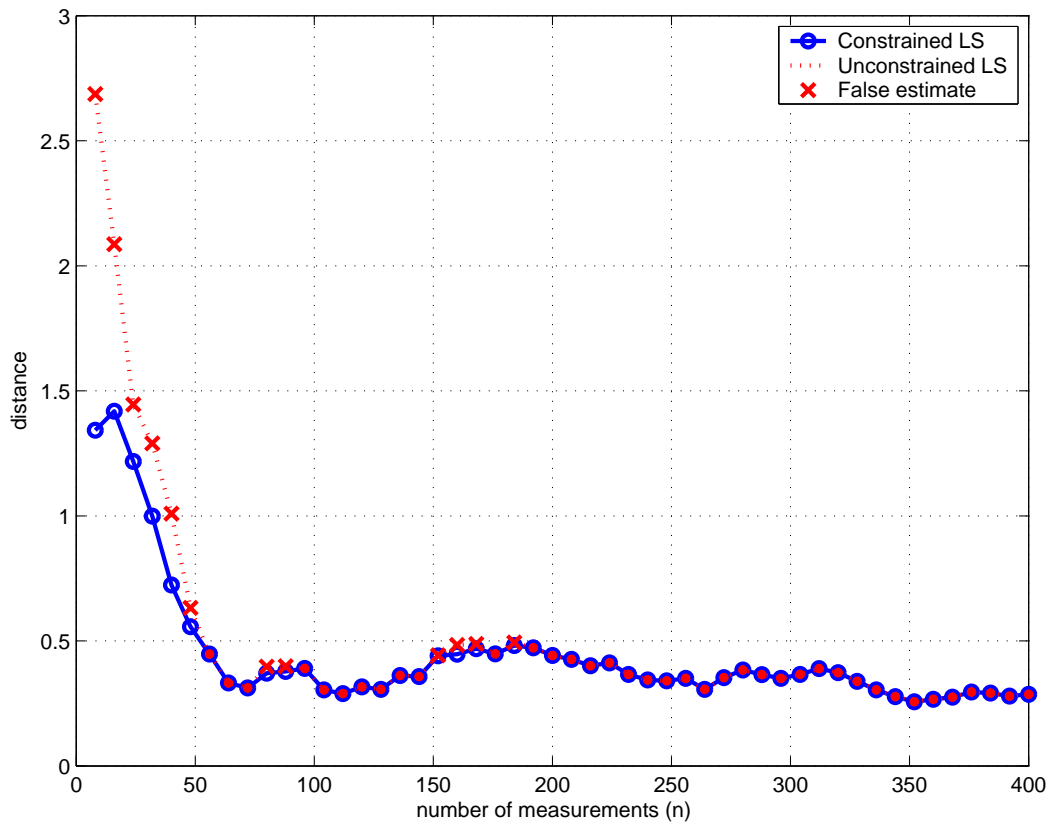}
\end{center}
\caption{\label{F:blochlength}
The Hilbert-Schmidt distance between the true $3 \times 3$ state
with eigenvalues $0.1186,\, 0.2871,\, 0.5943$ and the estimate. When the number
of the measurement is more than 200, the unconstrained estimate gives really
a positive semidefinite matrix.}
\end{figure}

\begin{figure}[!ht]
\begin{center}
\includegraphics[width=14cm]{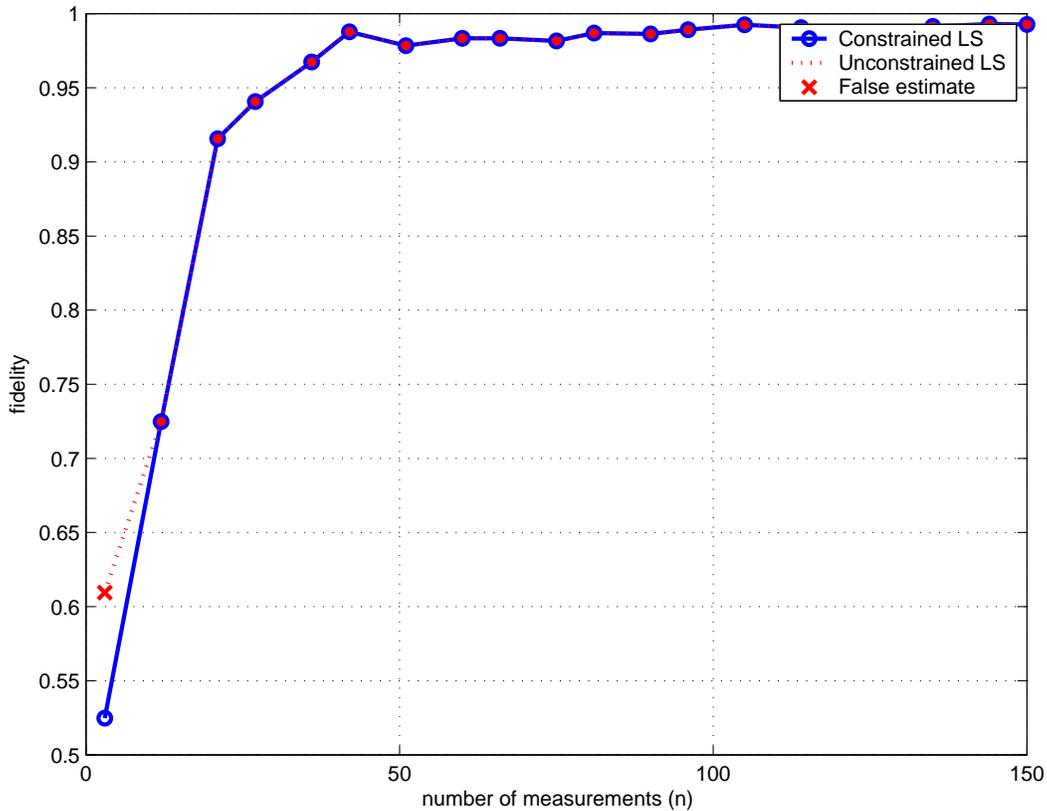}
\end{center}
\caption{\label{F:fid_mixed}
The fidelity between the true $2 \times 2$ mixed state and the
estimate. When the number of the measurement is more than 10, the
unconstrained and the constrained estimates are the same.}
\end{figure}

The properties of the unconstrained estimate $\Phi_n^{\un}$ depend very much on
the true state. If the eigenvalues of the true state are strictly positive
(and not very small), then the estimate is rather good and the convergence is
visible from the simulations, see Figure \ref{F:blochlength} and \ref{F:fid_mixed}.
The simulations are essentially simpler in the $2\times 2$ case, when the boundary
of the state space consists of pure states and the positivity of the estimate can be
seen from the length of the Bloch vector. In the $3\times 3$ case the boundary is more
complicated, it consists of the non-invertible densities.

\begin{figure}[!ht]
\begin{center}
\includegraphics[width=14cm]{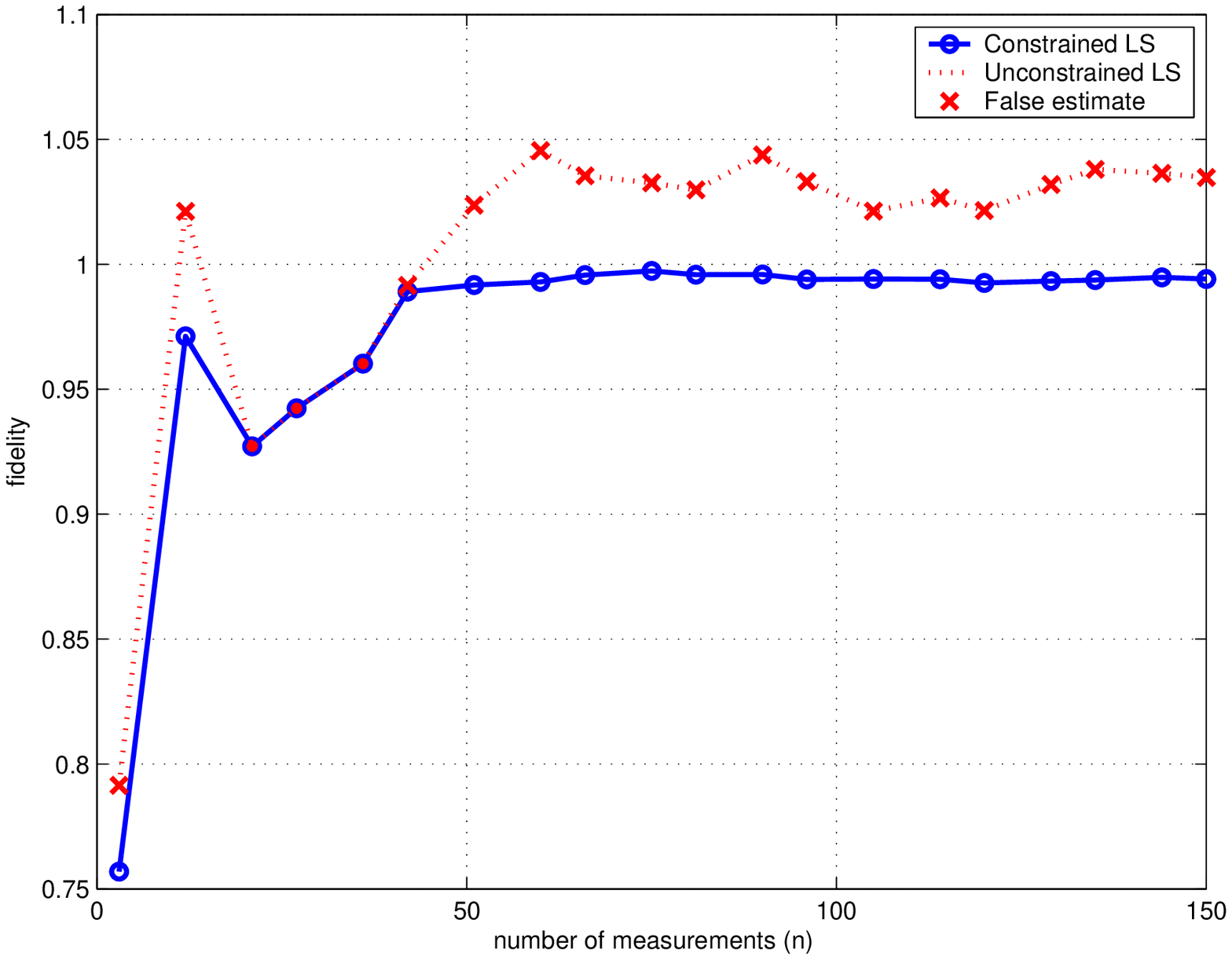}
\end{center}
\caption{The fidelity between the true $2 \times 2$ pure state and the
estimates. The unconstrained estimate is often outside of the Bloch ball
and in this case the (real part of the complex) fidelity can be bigger than 1. 
The constrained estimate converges to the true state.}
\end{figure}

\section{Constrained estimate}

There are cases when $\Phi_n^{\un}$ is not a positive semidefinite
matrix, sometimes we call  $\Phi_n^{\un}$ {\bf unconstrained estimate}.
The expectation value of $\Phi_n^{\un}$ is the true state of the system,
so it is an {\bf unbiased estimate}.

We can use the method of least squares to get a density matrix:
\begin{equation}\label{E:LQ}
\Phi_n :=\argmin_\omega \Tr (\Phi_n^{\un} -\omega)^2=\argmin_\omega \sum_{i,j}
(\Phi_n^{\un})_{ij}-\omega_{ij})^2\,,
\end{equation}
where $\omega$ runs over the density matrices. The density matrices form
a closed convex set $\iD_k$, therefore the minimizer is unique.
Note that for a qubit the closest positive semidefinite matrix is easy
to find. If the values of the estimates are simply the Bloch vectors,
then
\begin{equation} \label{E:qubit_constrained}
\Phi_n(x)=\cases{\Phi_n^{\un}(x) & if $\|\Phi_n^{\un}(x)\| \leq 1$, \cr
\phantom{mm}\cr
\displaystyle{\frac{\Phi_n^{\un}(x)}{\|\Phi_n^{\un}(x)\|}}\qquad & otherwise.}
\end{equation}

\begin{thm}
The constrained estimate $\Phi_n$ is asymptotically unbiased.
\end{thm}

\proof
We can use the fact that $\Phi_n^{\un}$ is unbiased and to show that
$\Phi_n$ is an asymptotically unbiased estimate we study their difference.
Let $p(x)$ be the probability of the measurement result $x$ and $X$ is the
set of outcomes such that $\Phi_n^{\un}(x)\ne \Phi_n(x)$, then evidently
\begin{equation}\label{10E:kint0}
\sum_{x} \Phi_n^{\un}(x) p(x)-\sum_{x} \Phi_n(x) p(x)=
\sum_{x\in X} (\Phi_n^{\un}(x)-\Phi_n(x)) p(x)\,.
\end{equation}
If $\iD_k\subset \iM_k$ is the set of density matrices, then $X$ is the set
of outcomes $x$ such that $\Phi_n^{\un}(x) \notin \iD_k$. Let us fix a norm
on the space $\iM_k$. (Note that all norms are equivalent.)

Let $\eps > 0$ be arbitrary. We split $X$ into two subsets:
$$
X_1=\{ x \in X:{\rm distance}(\Phi_n^{\un}(x),\iD_k)\le \eps\}
\quad \mbox{and}\quad X_2=X \setminus X_1.
$$
Note that ${\rm distance}(\Phi_n^{\un}(x),\iD_k)=\|\Phi_n^{\un}(x)-
\Phi_n (x)\|$. Then
$$
\sum_{x\in X} \|\tilde\Phi_n(x)-\Phi_n(x)\| p(x) \leq
\sum_{x\in X_1} \|\tilde\Phi_n(x)-\Phi_n(x)\| p(x)+
\sum_{x\in X_2} \|\tilde\Phi_n(x)-\Phi_n(x)\| p(x)\,.
$$
The first term is majorized by $\eps$ and the second one by  $C \Prob (X_1)$.
Since the first is arbitrary small and the latter goes to $0$,
we can conclude that (\ref{10E:kint0}) goes to $0$. \qed

\begin{figure}[!ht]
\begin{center}
\includegraphics[width=12cm]{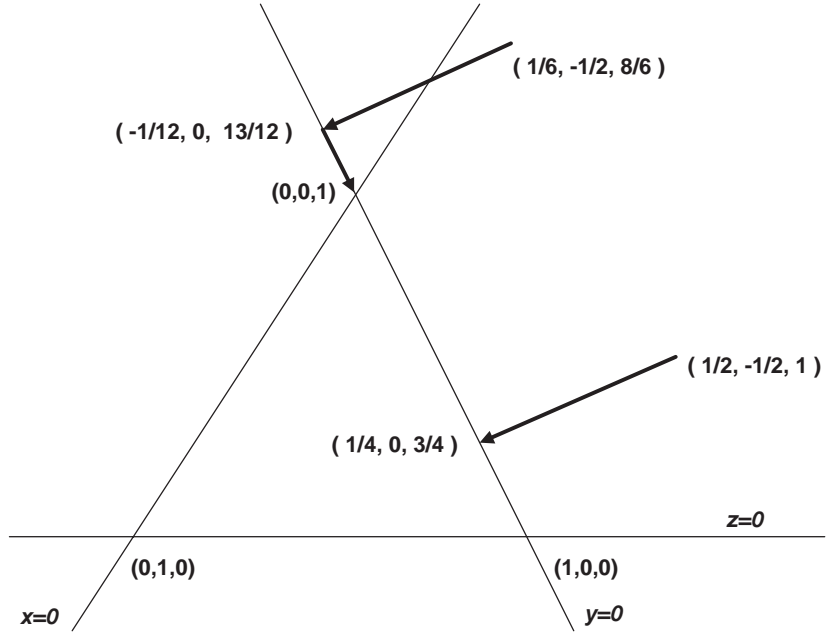}
\end{center}
\caption{\label{F:constrained}
The constrained estimate for $3\times 3$ matrices. The plain $x+y+z=1$
of $\bbbr^3$ is shown. The triangle $\{ (x,y,z): x,y,z \ge 0 \}$ corresponds
to the diagonal density matrices. Starting from the unconstrained estimate 
$\Diag (1/2,-1/2,1)$, the constrained $\Diag (1/4, 0, 3/4)$ is reached in one step.
Starting from $\Diag (1/6, -1/2, 8/6)$, two steps are needed.}
\end{figure}

{\bf Computing the constrained estimate.} The computation of the
minimizer of (\ref{E:LQ}) is easier if $\Phi_n^{\un}$ is diagonal,
assume that $\Phi_n^{\un}=\Diag (x_1, x_2, \dots, x_n)$ and
$x_1, x_2, \dots, x_k < 0$ and $x_{k+1}, x_{k+2}, \dots, x_n \ge 0$.
The minimizer is obviously diagonal, hence we need to solve
$$
\argmin_{y_i} \sum_{i} (x_{i}-y_{i})^2
$$
under the constraint $y_i \ge 0$ and $\sum_i y_i=1$. According to
the inequality between the quadratic and arithmetic means, we have
\begin{eqnarray*}
\sum_{i=1}^n (x_i-y_i)^2 &\ge& \sum_{i=1}^k x_i^2+\sum_{i=k+1}^n (x_i-y_i)^2 \ge 
\sum_{i=1}^k x_i^2+\frac{1}{n-k}\left(\sum_{i=k+1}^n (x_i-y_i)\right)^2 \cr
& = &\sum_{i=1}^k x_i^2+\frac{1}{n-k}\left(\sum_{i=1}^k  y_i-x_i \right)^2.
\end{eqnarray*}
If
$$
y_i=x_i+c \quad \left(i=k+1,k+2,\dots, n, 
\quad c=\frac{1}{n-k}\sum_{i=1}^k  x_i\right)
$$
are positive, then the minimizer is $(y_1,y_2,\dots, y_n)$, where
$y_1=y_2=\dots =y_k=0$ and the other $y_i$'s are defined above. If the
$n$-tuple $(y_1,y_2,\dots, y_n)$ contains negative entries, then we
repeat the procedure, the negative entries are replaced with 0 and
the actual value of $c$ is added to the other entries. After finitely
many steps we arrive at the minimizer. Figure \ref{F:constrained} shows 
the details for $n=3$. 

In the general case, we can change the basis such that $\Phi_n^{\un}$ becomes
diagonal, since the Hilbert-Schmidt distance is invariant under this 
transformation. So let $U\Phi_n^{\un}U^*=\Diag (x_1, x_2, \dots, x_n)$ 
for a unitary $U$. Then we compute the minimizer $\Diag(y_1, y_2, \dots, 
y_n)$ using the above procedure and
$$
\Phi_n = U^*\Diag(y_1, y_2, \dots, y_n)U\,.
$$

\section{Estimations for a qubit}

The mean quadratic error matrix may be used to measure the efficiency of an
estimate. If the unknown state is parametrized by $(\theta_1,\theta_2,
\dots,\theta_m)$, then the {\bf mean quadratic error} is an $n \times n$
matrix defined as
$$
V_n(\theta)_{i,j}:=\sum_{x \in \iX_n} (\Phi_n(x)_i -\theta_i)
( \Phi_n(x)_j -\theta_j)\,p(x) \qquad (1 \le i,j \le n).
$$
In case of a qubit, the Bloch parametrization can be used. Then $\theta=
(\theta_1,\theta_2,\theta_3)^t$ belongs to the unit ball of $\bbbr^3$.
($(\theta_1,\theta_2,\theta_3)^t$ means a column vector, so $^t$ may be regarded
as the transpose.)

\begin{pl}\label{pl:MUB}
Assume that the observables
$$
A(i)=\bu(i) \cdot \sigma \qquad (1 \le i \le 3)
$$
are measured in the true state
\begin{equation}\label{E:rhoth}
\rho_\theta=
\frac{1}{2}(I+\theta \cdot \sigma)=\frac{1}{2}\left[ \begin{array}{cc}
1+\theta_3& \theta_1 - \im \theta_2 \\ \theta_1 + \im \theta_2
& 1-\theta_3\end{array} \right],
\end{equation}
where $\bu(1), \bu(2)$ and $\bu(3)$ are unit vectors in $\bbbr^3$. The
spectral decomposition of $A(i)$ is
$$
1 \cdot \frac{1}{2}(I+ \bu(i) \cdot \sigma)+(-1)\cdot \frac{1}{2}(I-
\bu(i) \cdot \sigma)
$$
and
$$
p_i:=\Prob(A(i)= 1)=\frac{1 + \bu(i)\cdot\theta }{2}\,.
$$
If the measurements are performed $r$ times, then $\Prob(A(i)=1)$ is estimated by
the relative frequency $\nu(i)_r$ of the outcome $1$. The equations
$$
\nu(i)_r=\frac{1+ \bu(i)\cdot\hat \theta }{2} \qquad (1 \le i \le 3)
$$
should be solved to find an estimate. The solution is
\begin{equation}\label{E:sol}
\hat \theta=T^{-1}(\nu(1)_r,\nu(2)_r,\nu(3)_r)^t\,
\end{equation}
where $^t$ denotes the transpose (of a row vector) and
the matrix $T$ is
$$
T=\left[ \begin{array}{ccc} \bu(1)_1& \bu(1)_2 & \bu(1)_3  \\
\bu(2)_1& \bu(2)_2 & \bu(2)_3  \\
\bu(3)_1& \bu(3)_2 & \bu(3)_3 \end{array} \right].
$$
In particular, if each of the three measurements is performed once and
the result is $\eps=(\eps_1,\eps_2,\eps_3)^t$, then the unconstrained
estimate is
$$
\Phi^{\un}(\eps)= T^{-1}\eps .
$$
Similarly to (\ref{E:sol}), we have
\begin{equation}\label{E:sol2}
\theta=T^{-1}p\,.
\end{equation}
The  mean quadratic error matrix is the expectation of
$$
(T^{-1}\eps -\theta)(T^{-1}\eps -\theta)^t=T^{-1}\left((\eps-p)(\eps-p)^t
\right)(T^{-1})^*
$$
and the computation yields
\begin{equation}
V^{(1)}(\theta)= T^{-1}
\left[\begin{array}{ccc}1-(\bu(1)\cdot \theta)^2 & 0 & 0\\
0 & 1-(\bu(2)\cdot \theta)^2 & 0\\
0 & 0 & 1-(\bu(3)\cdot \theta)^2  \end{array}\right] (T^{-1})^*.
\end{equation}
When each measurement is performed $r$ times, then
$$
V_n^{(1)}(\theta)=\frac{1}{r}V^{(1)}(\theta),
$$
where $n=3r$.
If the observables $\sigma_1, \sigma_2$ and $\sigma_3$ are measured, then
\begin{equation}\label{E:comp}
V_n^{comp}(\theta)=\frac{1}{r}\left[\begin{array}{ccc}
1-\theta_1^2 & 0 & 0 \\ 0 & 1-\theta_2^2  & 0 \\
0 & 0 & 1-\theta_3^2
\end{array}\right].
\end{equation} \qed
\end{pl}

\begin{thm}
In the context of the previous example, the determinant of the average
mean quadratic error matrix is the smallest, if the vectors $\bu(1), \bu(2)$
and $\bu(3)$ are orthogonal, that is, the observables $A(1), A(2)$ and $A(3)$
are complementary.
\end{thm}

\proof
On the parameter space, Bloch ball, we consider the normalized Lebesgue measure.
(Any rotationally invariant measure may be considered and gives similar result.)
Since
\begin{eqnarray*}
\int V_n^{(1)}(\theta)\, d\theta&=&T^{-1}\Big(I-\int \Diag((\bu(1)\cdot \theta)^2,
(\bu(2)\cdot \theta)^2, (\bu(2)\cdot \theta)^2)d\theta\Big) (T^{-1})^*\cr
&=& C\, (T^*T)^{-1}
\end{eqnarray*}
with some positive constant $C$, the determinant is minimal if $\Det (T^*T)
=(\Det T)^2$ is maximal. $\Det\, T$ is the volume of the parallelepipedone
determined by the three vectors $\bu(1), \bu(2)$ and $\bu(3)$, and it is maximal
when they are orthogonal. \qed

The content of the theorem is similar to the result of \cite{WF}, however in the
approach of Wootters and Field not the mean quadratic error was minimized but the
information gain was maximized. The complementary (or unbiased) measurements are
optimal from both view point.

\begin{pl}
Let $\sigma_i=P_i-Q_i$ be the spectral decomposition and let
$$
F_i=\frac{P_i}{3} \quad \mbox{and} \quad F_{i+3}=\frac{Q_i}{3}
\qquad (1 \le i \le 3).
$$
be a POVM. The corresponding measurement is sometimes called {\bf standard qubit
tomography} \cite{REK} and it has 6 outcomes with probabilities
$$
p_i=\frac{1+\theta_i}{6}, \quad p_{i+3}=\frac{1-\theta_i}{6}
\qquad (1 \le i \le 3).
$$
The appropriate  (unconstrained) state  estimate
$$
\Phi(i)=\frac{1}{2}(I+3\sigma_i)=-I+3P_i, \quad
\Phi(i+3)=\frac{1}{2}(I-3\sigma_i)=-I+3Q_i\,.
$$
is unbiased $(1 \le i \le 3)$. If the true state is $\rho_\theta$ of (\ref{E:rhoth}),
then
$$
\sum_{j=1}^6 p_j \Phi(j)=\rho_\theta\,.
$$

The quadratic error matrix for $n$ independent measurements is
\begin{equation}\label{E:standard}
V_n^{stand}(\theta)=\frac{1}{n}\left[\begin{array}{ccc}
3-\theta_1^2 &
-\theta_1\theta_2 &
-\theta_1\theta_3\\
-\theta_1\theta_2 & 3-\theta_2^2 &
-\theta_2\theta_3\\
-\theta_1\theta_3 & -\theta_2\theta_3 &
3-\theta_3^2\end{array}\right].
\end{equation}
\end{pl}

\begin{prop}
In the context of the previous example, the complementary measurement is
more efficient than the standard one, i.e. its mean quadratic error matrix
is smaller.
\end{prop}

\proof
To compare the efficiency of the standard measurement and the complementary
measurement, we study the mean quadratic error matrices (\ref{E:comp}) and
(\ref{E:standard}). The difference $V_{n}^{stand}(\theta)-V_{n}(\theta)$
has the form
\begin{eqnarray*}
\frac1n\left[\begin{array}{ccc}3-\theta_1^2 & -\theta_1\theta_2 &
-\theta_1\theta_3\\
-\theta_1\theta_2 & 3-\theta_2^2 &
-\theta_2\theta_3\\
-\theta_1\theta_3 & -\theta_2\theta_3 &
3-\theta_3^2\end{array}\right]-\frac3n\left[\begin{array}{ccc}1-\theta_1^2
& 0 & 0\\0 & 1-\theta_2^2 & 0\\0 & 0 &
1-\theta_3^2\end{array}\right]\\\\ =
\frac1n \left[\begin{array}{ccc}2\theta_1^2 & -\theta_1\theta_2 &
-\theta_1\theta_3\\
-\theta_1\theta_2 & 2\theta_2^2 &
-\theta_2\theta_3\\
-\theta_1\theta_3 & -\theta_2\theta_3 &
2\theta_3^2\end{array}\right]=\frac1n\, \left[\begin{array}{ccc}2
& -1&
-1\\-1&2&-1\\-1&-1&2\end{array}\right]\circ\left(\left[\begin{array}{c}\theta_1\\
\theta_2\\
\theta_3\end{array}\right]\cdot \left[\begin{array}{ccc}\theta_1&
\theta_2& \theta_3\end{array}\right]\right)
\end{eqnarray*}
where $\circ$ stands for the Hadamard product. Since the Hadamard product of
two positive semidefinite matrices is positive semidefinite, we have
$V_{n}^{stand}(\theta) \ge V_{n}^{comp}(\theta)$. The complementary measurement
is  more effective, than the standard one. \qed

\begin{pl}\label{10pl:5}
Consider the following Bloch vectors
\begin{eqnarray*}
&&a_1 = \frac{1}{\sqrt{3}}(1,1,1), \qquad \,\,
a_2 = \frac{1}{\sqrt{3}}(1,-1,-1),\\
&&a_3 = \frac{1}{\sqrt{3}}(-1,1,-1), \quad
a_4 = \frac{1}{\sqrt{3}}(-1,-1,1) .
\end{eqnarray*}
and form the positive operators
\begin{equation} \label{E:minimalPOVM}
F_i=\frac{1}{4}(\sigma_0+a_i \cdot \sigma )\qquad (1 \leq i \leq 4).
\end{equation}
They determine a measurement, $\sum_{i=1}^4 F_i=I$. The probability of the
outcome $i$ is
$$
p_i=\Tr F_i \rho_\theta=\frac{1}{4}(1+ a_i \cdot \theta).
$$
The above POVM is called {\bf minimal qubit tomography} by Reh{\'a}cek, Englert 
and Kaszlikowski \cite{REK}.

The matrix-valued estimator 
$$
\Phi^{min}(i)=-\sigma_0+6 F_i \qquad (1 \leq i \leq 4).
$$
is unbiased. If the measurement is performed $n$ times, then 
the average (written in vector-valued form) is
\begin{equation}\label{minimal}
\Phi_{n}^{min}=3\sum_{i=1}^4\frac{n_i}{n} a_i\,
\end{equation}
where $n_i$ is the number of the outcome $i$ from the $n$ measurements.
The mean quadratic error matrix is
\begin{equation}\label{QE:minimal}
V_{n}^{min}(\theta)=\frac1n\left[\begin{array}{ccc}3-\theta_1^2 &
\sqrt{3}\theta_3-\theta_1\theta_2 &
\sqrt{3}\theta_2-\theta_1\theta_3\\
\sqrt{3}\theta_3-\theta_1\theta_2 & 3-\theta_2^2 &
\sqrt{3}\theta_1-\theta_2\theta_3\\
\sqrt{3}\theta_2-\theta_1\theta_3 &
\sqrt{3}\theta_1-\theta_2\theta_3 & 3-\theta_3^2\end{array}\right].
\end{equation}
Unfortunately, the above matrix is not comparable with the mean quadratic error
matrix (\ref{E:comp}), i.e. their difference is indefinite. However,
$\Tr V_{n}^{comp} \le \Tr V_{n}^{min}$.

\qed
\end{pl}

\section{Conclusion}

The estimation of the density matrix of a $k$-level quantum system
 is studied in this paper.
The essential ingredients of an estimation scheme are identified. Those
 are the parametrization of the density operator $\rho$,
 the observables to be measured, and the estimator mapping the measured values to
 an estimate of the density operator.
The considered parametrization is given by the real and imaginary part of
 the entries, and they are estimated by independent measurements.
A special set of commuting observables is defined in order to
 obtain measured values that are classical random variables.

The unconstrained estimate gives a matrix which may be not positive definite
and the constrained estimate is the closest density matrix with respect to
the Hilbert-Schmidt distance. The constrained estimate is given by a simple
procedure starting with the diagonalization of the unconstrained one.

It is established that the properties of the estimation procedure depend 
very much on the invertibility of the true state.
In case of an invertible true state, the unconstrained estimate becomes proper
relatively fast. It has been found that for pure states the unconstrained estimates,
that are self-adjoint by construction, may not be positive semidefinite and this
requires to apply a regularization called constrained estimation procedure. 

The estimation procedures carried out by different estimators are
compared based on the biasedness of the estimates and their mean quadratic
error matrices. In particular, several estimation schemes are compared for
the unknown state of a qubit when a single qubit is measured at a time, and
its density matrix is parametrized using the Bloch vector. It is shown that 
the average mean quadratic error matrix is the smallest if the applied 
observables are complementary.

The results are illustrated by computer simulations.

\end{document}